\documentclass[vecphys,graybox]{svmult}

\usepackage{mathptmx}       
\usepackage{helvet}         
\usepackage{courier}        
\usepackage{type1cm}        

\usepackage{makeidx}         
\usepackage{graphicx}        
\usepackage{multicol}        
\usepackage[bottom]{footmisc}
\usepackage{amsmath}
\usepackage{amsfonts}
\usepackage{amssymb}
\usepackage{booktabs} 

\usepackage{float}
\usepackage{bm}
\makeindex             

 \usepackage{hyperref}
\usepackage{xcolor}
\definecolor{dark-red}{rgb}{0.6,0.15,0.15}
\definecolor{dark-blue}{rgb}{0.15,0.15,0.8}
\definecolor{medium-blue}{rgb}{0,0,0.6}
\hypersetup{
    colorlinks, linkcolor={dark-red},
    citecolor={dark-blue}, urlcolor={medium-blue}
}
\begin{document}

\title*{Comparing results for a global metric from analytical perturbation theory and a numerical code}
\titlerunning{Comparing metrics from analytical perturbations and a numerical code} 
\author{J. E. Cuch\'i, A. Molina, E. Ruiz}
\institute{J. E. Cuch\'i \at Departamento de F\'isica Fundamental, Universidad de Salamanca \email{jecuchi@gmail.com}
\and A. Molina \at Departamento de F\'isica Fonamental, Universidad Aut\'onoma de Barcelona \email{alfred.molina@ub.edu}
\and E. Ruiz \at Departamento de F\'isica Fundamental, Universidad de Salamanca \email{eruiz@usal.es}
}
%
%
\maketitle

%
\abstract{We compare the results obtained from analytical perturbation theory and the AKM numerical code for an axistationary spacetime built from matching a rotating perfect fluid interior with the equation of state $\epsilon-3p=4B$ of the simple MIT bag model and an asymptotically flat exterior. We discuss the behaviour of the error in the metric components of the analytical approximation going to higher orders. Additionally, we check and comment the errors in multipole moments, central pressure and some other physical properties of the spacetime.}

\section{Introduction}

The lack of stellar  models in General Relativity --i.\,e. a stationary and axisymmetric perfect fluid interior matched with an asymptotically flat vacuum exterior-- is in direct contrast with the importance they could have for the astrophysics of compact stars and in particular the determination of their possible compositions. One of such possibilities is the interesting case of strange matter.
 A versatile model for strange matter is a simple MIT bag model with equation of state (EOS) $\epsilon-3p=4B$. We will study this stellar model with the results provided by the CMMR post-Minkowskian and slow rotation approximation scheme \cite{cabezas2007ags} and its behaviour when we go to higher orders of approximation. Also, we will give the relative error in these functions and quantities when compared with very precise numerical results obtained with the AKM code \cite{ansorg2002highly,ansorg2003highly}.

\section{The analytical and numerical metrics}
The stationary and axisymmetric spacetime we study is the following. The interior $\mathcal{V^-}$ is filled with a perfect fluid with the EOS $\epsilon-3p=\epsilon_0$ and in ridig rotation so that, if $\bm{\xi},\,\bm{\chi}$ are the Killing vectors, its velocity is $\bm{u}=\psi (\bm{\xi}+\omega \bm{\chi})$, with $\psi$ a normalization factor and $\omega$ a constant correspondig to the rotation speed of the fluid as seen by a distant observer. The exterior $\mathcal{V^+}$ is asymptotically flat vacuum and is matched with the interior imposing continuity of the metrics and their first derivatives on the $p=0$ surface.

In CMMR, we solve the Einstein equations using a truncated multipolar post-Minkowskian approximation in spherical-like coordinates associated to harmonic ones. The post-Minkowskian parameter is $\lambda=m/r_s$, whith $m$ the Newtonian mass of the source and $r_s$ the coordinate radius of the static fluid; a different parameter $\Omega^2=\omega^2 r_s^3/m$ (the ratio between Newtonian centrifugal and gravitational forces), gives the truncation point of the  expansion in spherical harmonics, in this case preserving terms up to $\mathcal{O}(\Omega^3)$. The metric in each spacetime  $\bm{g}^\pm(\lambda,\,\Omega)$ is decomposed in Minkowski $\bm{\eta}$ plus the deviation $\bm{h}^\pm(\lambda,\,\Omega)$ and then Einstein's equations can be solved iteratively. The spacetimes are matched on the $p=0$ surface which can be expanded as $r_\Sigma=r_s\left[1+\sigma \Omega^2 P_2(\cos \theta)\right]+\mathcal{O}(\Omega^4)$ with $\sigma$ constant. Finally, the global metric depends only on $\epsilon_0,\, \omega$ and $r_s$. In \
\cite{Cuchi:2012nmNOREVTEX} (CGMR) we obtained the $\mathcal{O}(\lambda^{5/2},\,\Omega^3)$ metric for the EOS $\epsilon+(1-k)p=\epsilon_0$. Here we use its $(k=4,\,\epsilon_0=4B)$ subcase corresponding to the simple MIT bag model but now including terms up to  $\mathcal{O}(\lambda^{9/2},\,\Omega^3)$.

\begin{figure}[b]
\begin{tabular}{cc}
 \includegraphics[scale=0.095]{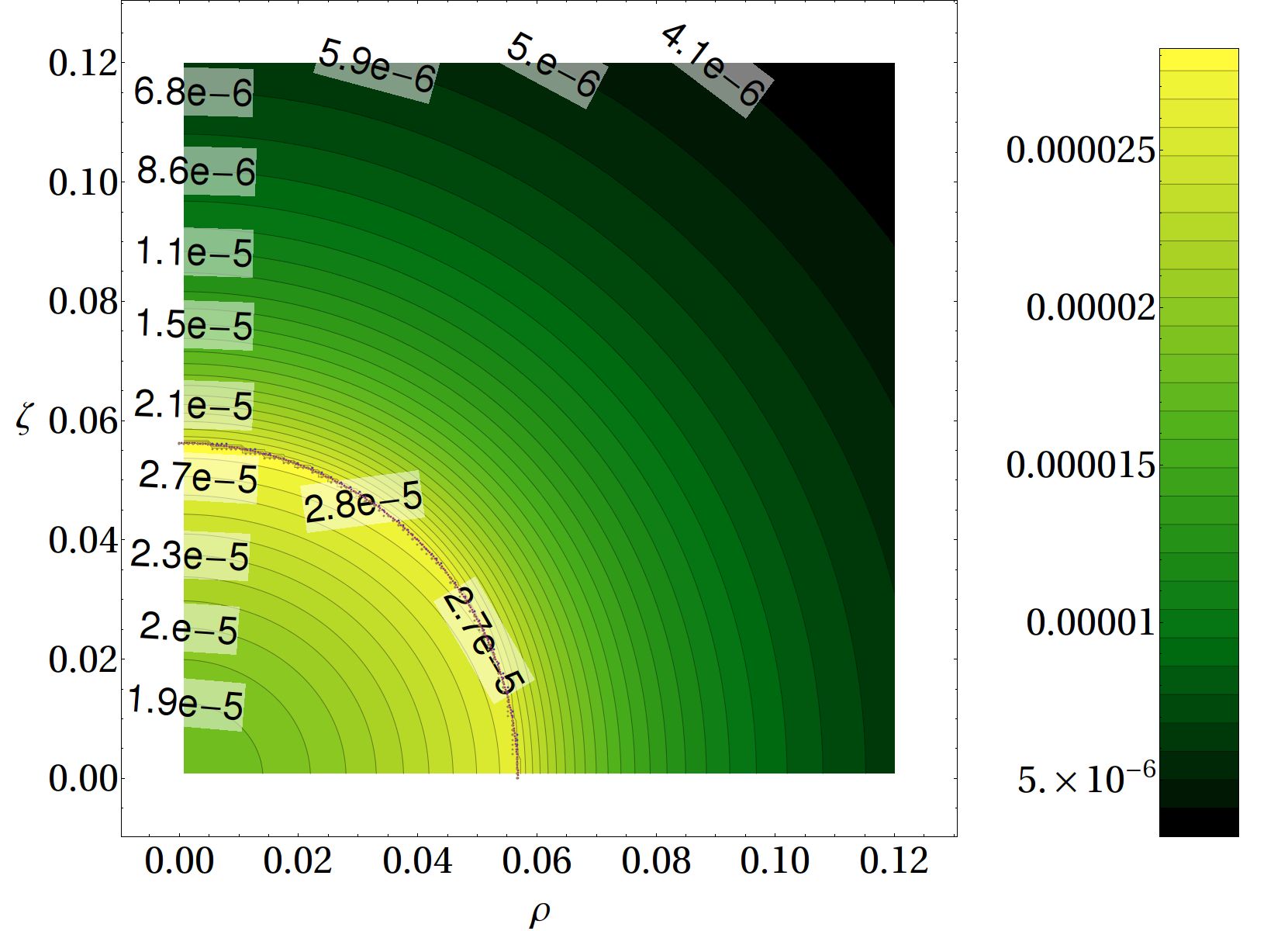}&
 \includegraphics[scale=0.095]{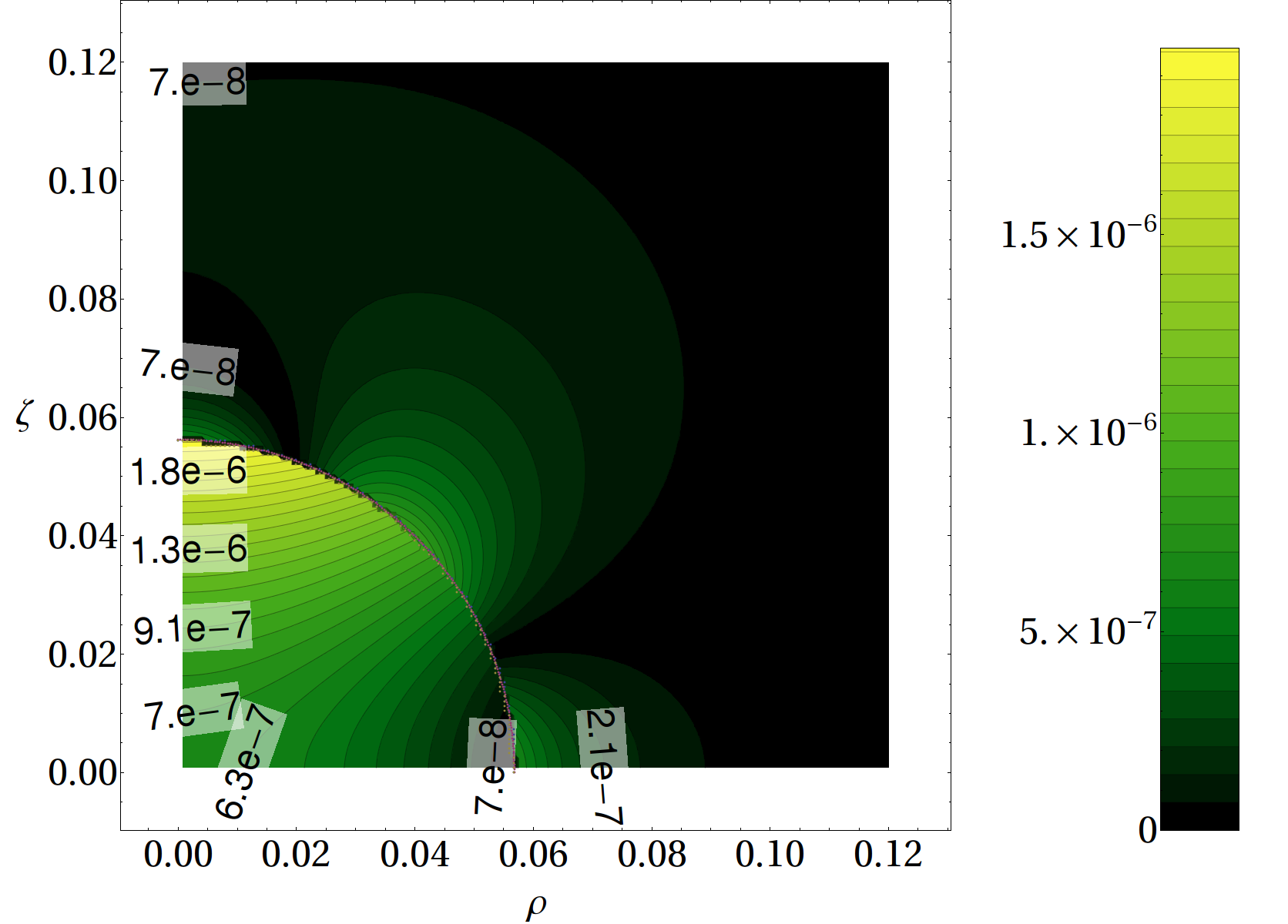}\\[-2ex]
\begin{small}(a)\end{small}&\begin{small}(b)\end{small}
\end{tabular}
\caption{Relative error for constant density in $g_{tt}$ using $M_0=8\times 10^{-4},\,\omega=0.2$ using CMMR up to: (a)  $\mathcal{O}(\lambda^{5/2},\,\Omega^3)$; (b) $\mathcal{O}(\lambda^{9/2},\,\Omega^3)$. The thin dotted lines represent the AKM and CMMR surfaces (indistinguishable in this picture size)}
\label{ere2012f1}
\end{figure}

AKM is a multi-domain spectral code that gives the matched metric on a grid over a quadrant of finite size. The grid coordinates are $\{\rho,\zeta\}$ , cylindrical associated to  quasi-isotropic coordinates and the resolution is customizable. It also gives some physical properties, like the first mass and angular momentum multipole moments $M_0$ and $J_1$, baryon rest mass $M_\text{b}$, circumpherential radius $R_\text{circ}$, binding energy $E_\text{b}$, polar and equatorial coordinate radii $r_\text{p},\,r_\text{e}$, polar redshift $z_\text{p}$ and central pressure and specific enthalpy $p_\text{c},\,h_\text{c}$. To build a stellar model, it needs goal values for any pair of these quantities as well as to fix some parameters to specify the EOS and provide an initial data file. The code is able to obtain initial data for different EOS if one manages to avoid unphysical configurations in the process. The precision it 
gets depends on the number $n$ of Chebyshev polynomials used, and can reach machine accuracy for high enough $n$ when the deformation of the source is not extreme. We fix it here to $n=20$.

%
%
%
\begin{figure}[bt]
\begin{tabular}{cc}
\includegraphics[scale=0.095]{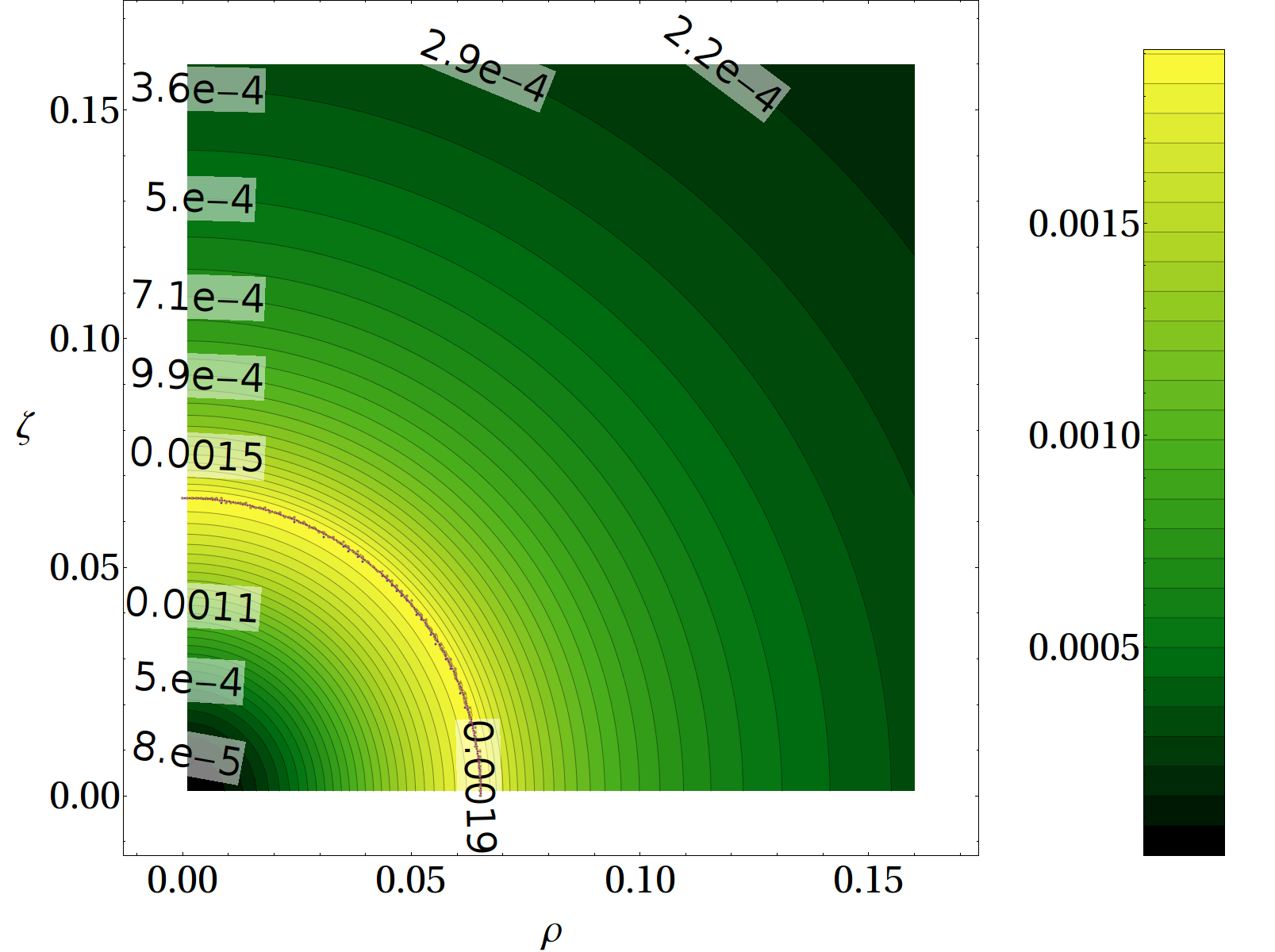}&
\includegraphics[scale=0.095]{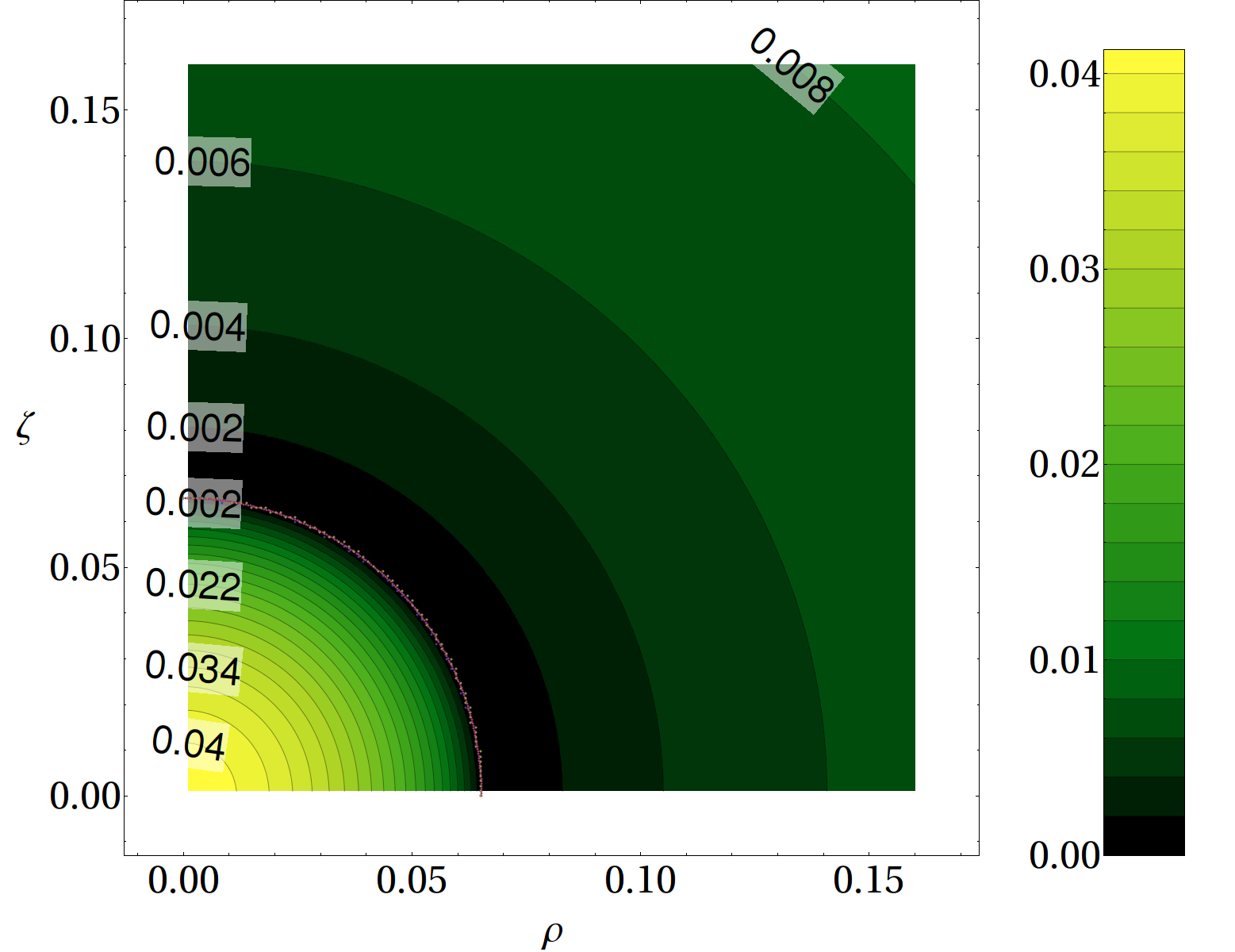}\\[-2ex]
\begin{small}(a)\end{small}&\begin{small}(b)\end{small}\\
\includegraphics[scale=0.095]{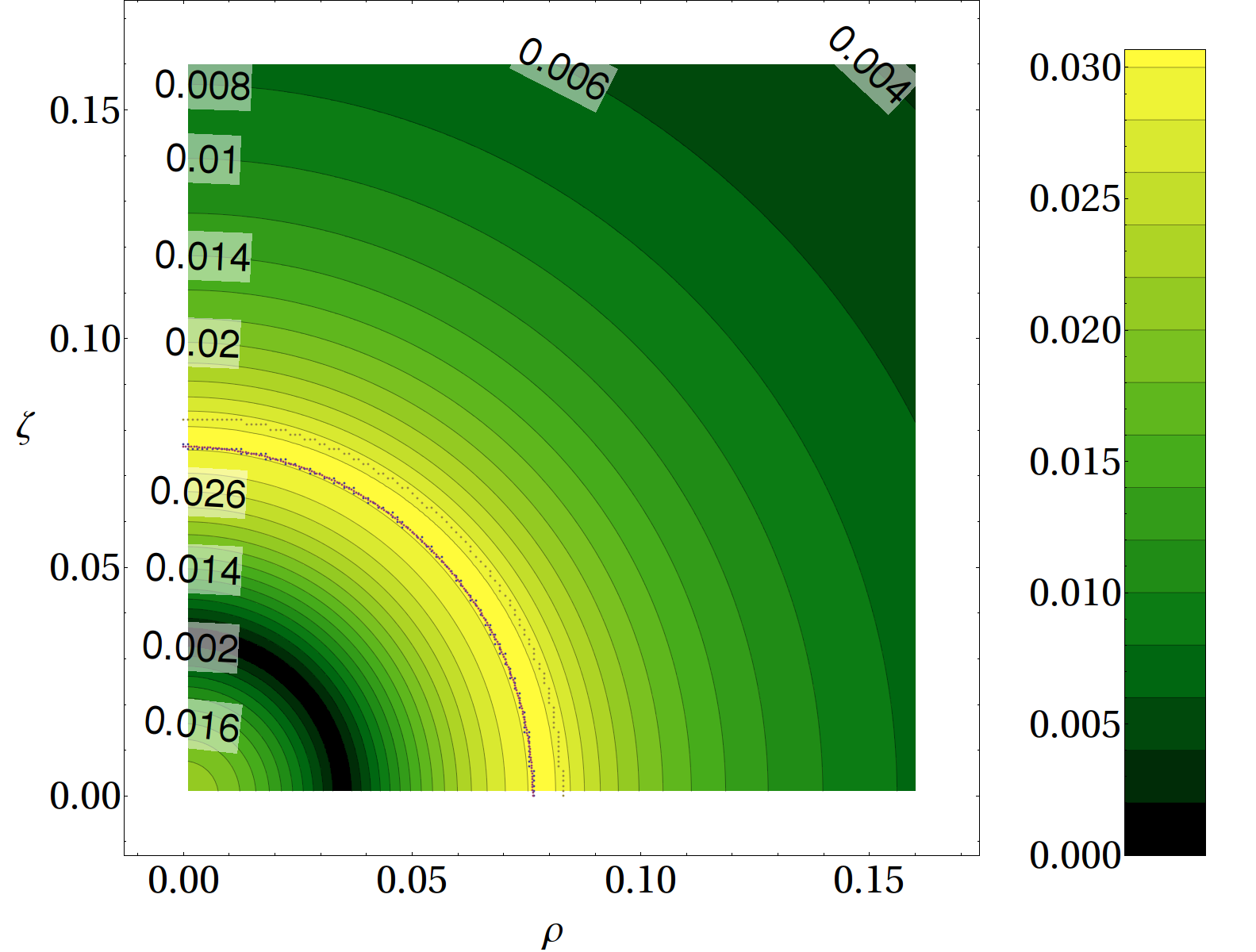}&
\includegraphics[scale=0.095]{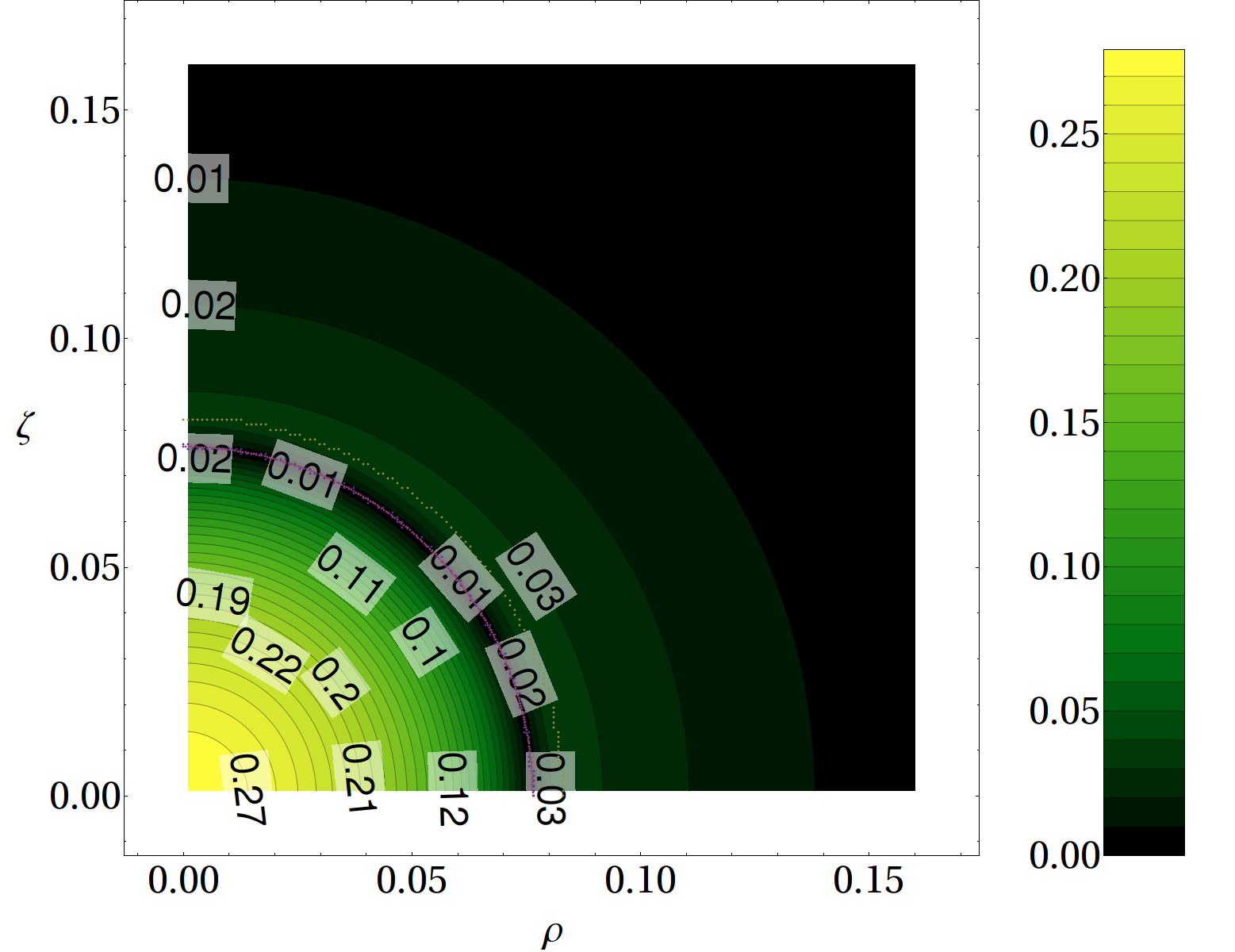}\\[-2ex]
\begin{small}(c)\end{small}&\begin{small}(d)\end{small}
\end{tabular}
\caption{Relative error between $\mathcal{O}(\lambda^{9/2},\,\Omega^3)$ CMMR and $n=20$ AKM in (a-b) $g_{tt}$ and $g_{t\varphi}$ for $M_0=8\times10^{-3}$, $\omega=0.24$ $(\lambda\approx0.077,\,\Omega\approx0.059)$; (c-d) $g_{tt}$ and $g_{t\varphi}$ for $\smash{M_0=0.0184}$, $\omega=0.24$ $(\lambda\approx0.12,\,\Omega\approx0.059)$. The thin dotted lines represent the AKM and CMMR surfaces.}
\label{ere2012f2}
\end{figure}
\section{Comparison and results}

 We build stellar models for different values of $(M_\text{b},\,\omega)$ and compare the metrics on the cylindrical-like coordinate grid of AKM. Working in units where $(c=G=B=1)$, the $k=4$ CGMR has only $(r_s,\,\omega)$ as free parameters. Unlike $\omega$, $r_s$ has no equivalent in AKM, so we must adjust its value. Hence, for each model we take the value of one of the AKM physical quantities and equating it to its CMMR counterpart we get $r_s$. Using $M_0,\,p_\text{c},$ and $J_1$ for this gives similar results. Here we use $M_0$. 

The results of Fig. \ref{ere2012f2} show the relative error in the metric $(g_{tt},g_{t\phi})$ components for $M_0=0.51\,M_\odot$ and the typical $M_0=1.39\,M_\odot$ for a moderately fast rotation frequency $\nu=102\,$ Hz near the source. The first case have a reasonable error (higher in $g_{t\phi}$) and the second must be improved. Going to higher orders in the $\lambda$ approximation is now automatic with the help of our \textit{Mathematica} subroutines, but improving the slow-rotation approximation is more cumbersome. Nevertheless, Fig. \ref{ere2012f1} shows the improvement for a $\epsilon=\epsilon_0$ EOS of moving from $\mathcal{O}(\lambda^{5/2},\,\Omega^3)$ to $\mathcal{O}(\lambda^{9/2},\,\Omega^3)$. It make us confident that, since our current error graphics do not show the lobular aspect of Fig. \ref{ere2012f1}(b), which means that the truncation of the multipolar expansion is an important error source, our results can be improved very easily even in the case of the strong gravitational field of a 
compact source of realistic mass.

\begin{table}
\caption{Some CMMR values and relative error with respect to AKM in $(c=G=B=1)$ units. In other units, the three models rotate at $\nu=102\,$Hz and their masses in $M_\odot$ appear in the first row.}
%
%
\begin{tabular}{p{1cm}p{1.6cm}p{1.7cm}p{1.6cm}p{1.7cm}p{1.6cm}p{1.7cm}}
\hline\noalign{\smallskip}
$M_\odot$&\multicolumn{2}{c}{$M_0=0.054,\ M_\text{b}=0.06$} &\multicolumn{2}{c}{$M_0=0.51,\ M_\text{b}=0.6$}	&\multicolumn{2}{c}{$M_0=1.39,\  M_\text{b}=1.7  $}	\\
\noalign{\smallskip}\svhline\noalign{\smallskip}
		&\text{CMMR}	&error	&\text{CMMR}	&error	&\text{CMMR}	&error\\
\midrule
$\omega$			&0.24	&			&0.24		&		&0.24		&\\
$M_\text{b}$			& 7.9997e-4	&3.2e-5		&7.95e-3	&6.1e-3		&0.02241	&0.0406\\
$M_0$				&7.104e-4	&		&6.772e-3	&		&0.0184		&\\
$J_1$				&8.3609e-8	&1e-4		&3.586e-6	&0.012		&1.82e-5	&0.034\\
$R_\text{circ}$			&0.034769	&3.6e-5		&0.07287	&8.6e-3		&0.102		&0.068\\
$E_\text{bind}$			&8.8730e-6	&2.8e-4		&3.766e-4	&0.040		&1.75e-3	&0.19\\
$z_\text{p}$			&0.021133	&7.6e-5		&0.1079		&0.016		&0.24		&0.13\\
$V_0$				&-0.020913	&4.6e-5		&-0.103		&0.010		&-0.221		&0.087\\
$r_\text{p}$			&0.033914	&4.7e-5		&0.0657		&9.8e-3		&0.0831		&0.09\\
$r_\text{e}$			&0.034055	&3.8e-5		&0.06593	&9.8e-3		&0.0834		&0.091\\
$r_\text{ratio}$		&0.995862	&9.4e-6		&0.9964		&2.4e-5		&0.997		&3e-4\\
$\epsilon_\text{c}$		&4.5935		&6.8e-5		&4.712		&0.019		&5.56		&0.21\\
$p_\text{c}$ 			&0.043679	&2.4-4		&0.254		&0.052		&0.631		&0.37\\
$h_\text{c}$			&0.908786	&1.7e-6		&0.9522		&2e-3		&1.025		&0.042\\
\noalign{\smallskip}\hline\noalign{\smallskip}
\end{tabular}
\label{ere2012t1}
\end{table}

\begin{acknowledgement}
 JEC thanks Junta de Castilla y Le\'on for grant EDU/1165/2007.  This work was supported by grant FIS2009-07238 (MICINN).
\end{acknowledgement}


%
 \bibliography{bibliografMACROsepnames}

\begin{thebibliography}{1}
\providecommand{\url}[1]{\texttt{#1}}
\providecommand{\urlprefix}{URL }
\expandafter\ifx\csname urlstyle\endcsname\relax
  \providecommand{\doi}[1]{doi:\discretionary{}{}{}#1}\else
  \providecommand{\doi}{doi:\discretionary{}{}{}\begingroup
  \urlstyle{rm}\Url}\fi
\providecommand{\eprint}[2][]{\url{#2}}

\bibitem{ansorg2002highly}
Ansorg, M., Kleinw{\"a}chter, A., Meinel, R.
\newblock Highly accurate calculation of rotating neutron stars.
\newblock Astron. \& Astrophys. \textbf{381}(3), L49--L52 (2002)

\bibitem{ansorg2003highly}
Ansorg, M., Kleinw{\"a}chter, A., Meinel, R.
\newblock Highly accurate calculation of rotating neutron stars. {D}etailed
  description of the numerical methods.
\newblock Astron. \& Astrophys. \textbf{405}, 711--721 (2003)

\bibitem{cabezas2007ags}
Cabezas, J.A., Mart{\'\i}n, J., Molina, A., Ruiz, E.
\newblock {An approximate global solution of Einstein's equations for a
  rotating finite body}.
\newblock Gen. Relativ. Gravit. \textbf{39}(6), 707--736 (2007)

\bibitem{Cuchi:2012nmNOREVTEX}
Cuch{\'\i}, J.E., Gil-Rivero, A., Molina, A., Ruiz, E.
\newblock An approximate global solution of {E}instein's equations for a
  rotating compact source with linear equation of state  (2012).
\newblock Arxiv preprint 1212.4456 submitted to Gen. Rel. Grav.

\end{thebibliography}
\end{document}